\documentclass[preprint]{aastex631}
\usepackage{natbib}
\usepackage{enumitem}
\usepackage{placeins}
\usepackage{amsmath}
\usepackage{xcolor}

\definecolor{color1_sron_hc}{RGB}{221,170,51}
\definecolor{color2_sron_hc}{RGB}{187,85,102}
\definecolor{color3_sron_hc}{RGB}{0,68,136}
\definecolor{color1_sron_bright}{RGB}{68,119,170}
\definecolor{color2_sron_bright}{RGB}{102,204,238}
\definecolor{color3_sron_bright}{RGB}{34,136,51}
\definecolor{color4_sron_bright}{RGB}{204,187,68}
\definecolor{color1_sron_vibrant}{RGB}{0,119,187}
\definecolor{color2_sron_vibrant}{RGB}{51,187,238}
\definecolor{color3_sron_vibrant}{RGB}{0,153,136}
\definecolor{color4_sron_vibrant}{RGB}{238,119,51}

\begin{document}

\shorttitle{ALMA Proposal Assignment}
\shortauthors{Carpenter et al.}

\title{Enhancing Peer Review in Astronomy:\\ A Machine Learning and Optimization Approach to Reviewer Assignments for ALMA}

\author[0000-0003-2251-0602]{John M. Carpenter}
\affiliation{Joint ALMA Observatory, Avenida Alonso de C\'ordova 3107, Vitacura, Santiago, Chile}

\author{Andrea Corvill\'on}
\affiliation{Joint ALMA Observatory, Avenida Alonso de C\'ordova 3107, Vitacura, Santiago, Chile}

\author[0000-0001-5158-9677]{Nihar B. Shah}
\affiliation{Machine Learning and Computer Science Departments, Carnegie Mellon University, Pittsburgh, PA 15213, USA}

\correspondingauthor{J. Carpenter}
\email{john.carpenter@alma.cl}

\begin{abstract}
The increasing volume of papers and proposals that undergo peer review emphasizes the pressing need for greater automation to effectively manage the growing scale. In this study, we present the deployment and evaluation of machine learning and optimization techniques to assign proposals to reviewers that were developed for the Atacama Large Millimeter/submillimeter Array (ALMA) during the Cycle 10 Call for Proposals issued in 2023.  Using topic modeling algorithms, we identify the proposal topics and assess reviewers' expertise based on their previous ALMA proposal submissions. We then apply an adapted version of the assignment optimization algorithm from PeerReview4All \citep{Stelmakh21} to maximize the alignment between proposal topics and reviewer expertise. Our evaluation shows a significant improvement in  matching reviewer expertise: the median similarity score between the proposal topic and reviewer expertise increased by 51 percentage points compared to the previous cycle, and the percentage of reviewers reporting expertise in their assigned proposals rose by 20 percentage points. Furthermore, the assignment process proved highly effective in that no proposals required reassignment due to significant mismatches, resulting in a savings of 3 to 5 days of manual effort.
\end{abstract}

\section{Introduction}
\label{sec:intro}

Peer review is the cornerstone of modern scientific research, playing a crucial role in the allocation of billions in grant funding and providing access to high-cost facilities such as supercomputers and telescopes, both ground- and space-based. Traditionally, this process relies on an invited panel of experts to evaluate and recommend which proposals should be accepted. This model has been widely adopted across the scientific community, including most major astronomical observatories. A prominent example is the Atacama Large Millimeter/submillimeter Array (ALMA), which implemented panel-based peer review when it began science operations in 2011. As the largest ground-based observatory and the most sensitive telescope ever constructed for high-resolution imaging of the submillimeter sky, ALMA's high demand necessitates an efficient peer review process.

The number of proposals submitted to ALMA has steadily increased, from 919 in 2011 (Cycle 0) to a peak of 1836 in 2018 (Cycle 6). Currently, approximately 1700 proposals are submitted annually. This growing volume created logistical challenges for the traditional panel-based process, prompting ALMA to adopt a distributed peer review system. In this system, each proposal team designates one member to review ten proposals. ALMA first implemented distributed peer review in 2019 (Cycle 7) for a "supplemental" call to fill unallocated time on the Morita Array \citep{Carpenter20b}. Distributed peer review was then used for a subset of proposals in the main call in 2021 (Cycle 8) and fully adopted for all proposals, except Large Programs, in 2022 (Cycle 9). A comprehensive description of ALMA's distributed peer review process is provided by \citet{DonovanMeyer22}.

While distributed peer review addressed logistics challenges, it introduced complexities in assigning proposals to reviewers. Initially, assignments were based on common categories and keywords between the submitted proposals and the expertise self-identified by the reviewers. The top priority was to assign proposals where one or more of the proposal keywords specified by the principal investigator (PI) overlapped with the expertise keywords specified by the reviewer. Proposals that had the fewest available expert reviewers were assigned first. This process successfully aligned proposal and reviewer keywords in about 95\% of cases. To gauge the success of the assignment process from the reviewer's perspective, each reviewer had the option to rate their expertise on their assignments, with possible options of ``My field of expertise'', ``Some knowledge'', or ``Little or no knowledge''. Reviewers reported at least some knowledge in 90\% of the assigned proposals, with only 10\% categorized as ``little or no knowledge" \citep{DonovanMeyer22}. 

Despite its successes, the keyword-based approach faced limitations. In the 10\% of cases where reviewers reported little or no expertise, the assigned proposal keywords often still matched the reviewer’s expertise \citep{DonovanMeyer22}. This discrepancy could result from broad or imprecise keywords, conservative self-assessments of expertise by the reviewers, or inexperience among reviewers, particularly student participants. Additionally, the keywords provided by reviewers may not fully capture their expertise, complicating the identification of suitable matches. Another challenge is the logistical need to assign 10 proposals to each reviewer and ensure each proposal is reviewed 10 times. In some instances, reviewers received proposals outside their primary field (e.g., a circumstellar disk proposal assigned to an extragalactic expert). Resolving these mismatches required three to five days of manual effort to swap assignments, making it the most time-consuming step in the assignment process.

Other observatories and scientific disciplines have experienced similar challenges and have turned to advanced machine learning and optimization techniques to enhance the efficiency of proposal assignments. These methods typically operate in two phases~\citep[see Section 3 in][]{Shah22}: (1) machine learning methods quantify the similarity between a proposal's subject matter and a reviewer's expertise, and (2) optimization algorithms assign proposals to reviewers based on a defined metric.

For example, the Space Telescope Science Institute (STScI) employs machine learning techniques to classify Hubble Space Telescope (HST) and James Webb Space Telescope (JWST) proposals into broad topics. Reviewer expertise is inferred from their publication abstracts in the Astrophysics Data System (ADS),\footnote{\url{https://ui.adsabs.harvard.edu}} enabling effective reviewer-proposal matching \citep{Strolger17, Strolger23}. Similarly, the European Southern Observatory (ESO) implemented machine learning techniques for distributed peer review \citep{Kerzendorf19, Kerzendorf20, Jerabkova23}. Proposal topics are inferred from proposal text, while reviewer expertise is determined from publication histories in ADS.

In computer science, machine learning and optimization techniques have long been integral to assigning papers submitted to conferences during distributed peer review. Major conferences, such as NeurIPS, handle over 13,000 submissions~\citep{NeurIPSfactsheet23}, necessitating automated tools. These tools compute similarity scores using methods such as automated text matching~\citep{charlin12framework, mimno07topicbased, cohan2020specter}, intersection of reviewer- and author-selected subject areas, and manual reviewer ``bidding" to express preferences~\citep{fiez2020super}. Optimization algorithms then assign reviewers to papers, maximizing similarity scores or addressing fairness concerns~\citep{Stelmakh21, kobren19localfairness, stelmakh2023gold}.

Inspired by the efforts at other observatories and in computer science, ALMA developed a new proposal assignment approach with three primary objectives. First, the proposal assignment process should result in an overall increase in the expertise as measured by the similarity between the proposal topic and the reviewer's expertise. Second, reviewers should indicate a higher level of expertise in their assignments. Finally, the process needs to be automated and robust to reduce the amount of time spent on manual reassignments. 

This paper presents the implementation and evaluation of the new assignment system. Section~\ref{sec:implementation} describes the methodology for computing similarities between a reviewer's expertise and the submitted proposals, as well as optimizing assignments using an adapted version of the PeerReview4All algorithm \citep{Stelmakh21}. Section~\ref{sec:evaluation} establishes the utility of the similarity measures by correlating Cycle 8 and Cycle 9 assignments with reviewers' self-reported expertise and evaluates the performance of the overall assignment process following its full deployment in Cycle 10. 
Section~\ref{sec:summary} provides a summary of our findings and suggestions for future improvements.

\section{Implementation}
\label{sec:implementation}

\begin{figure}[ht]
\centering
\includegraphics{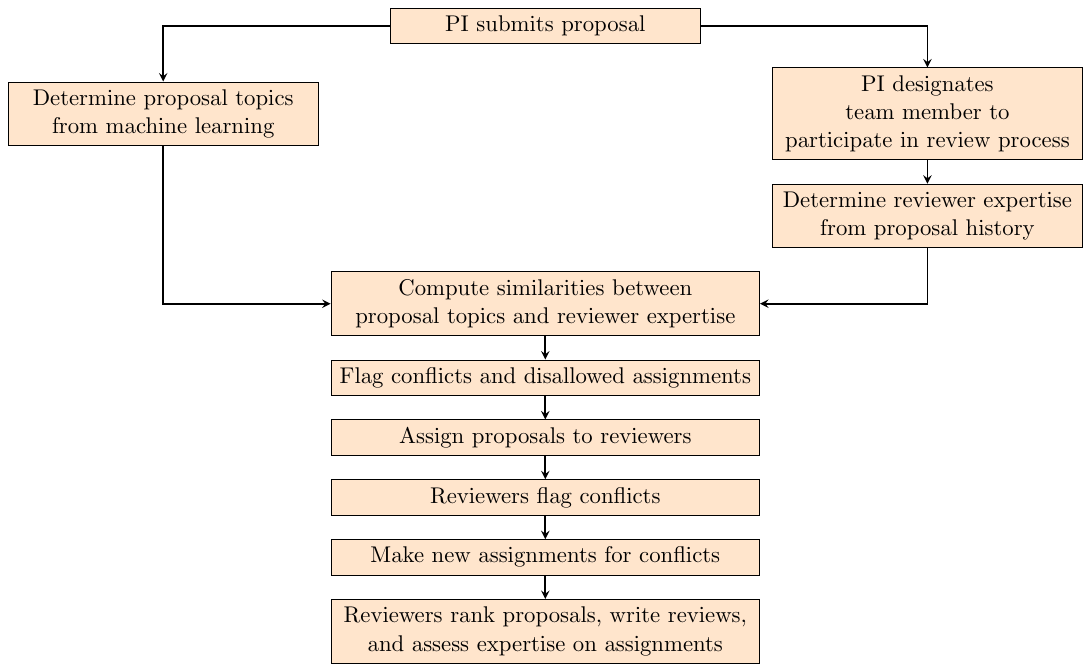}
\caption{
  \label{fig:flowchart}
  Flowchart of the proposal assignment process used by ALMA in Cycle 10, starting from the proposal submission by the PI and ending with the reviewer providing their rankings and self-assessment of their expertise on individual assignments.
}
\end{figure}

This section describes the deployment of the proposal assignment process adopted by ALMA and implemented for 2023 (Cycle 10). Figure~\ref{fig:flowchart} provides an overview of the process from the proposal submission by the PI to the ranking of proposals by a reviewer. The submitted proposal is passed to unsupervised machine-learning algorithms to infer the topics of the proposal (see Section~\ref{subsec:proptopics}). The PI designates a reviewer that will participate in the review process, where the scientific expertise of the reviewer is established from their proposal history, also based on machine learning techniques (see Section~\ref{subsec:expertise}). The similarities between the proposal topics and the reviewer's expertise is computed, and after flagging conflicts and disallowed proposal assignments, the proposals are assigned to the reviewers (see Section~\ref{subsec:assignments}). The following subsections describe the deployment in more detail.

\subsection{Topic Modeling}
\label{subsec:proptopics}

The first step in the proposal assignment process is to determine the topics of each proposal using natural language processing. The PDF files for submitted ALMA proposals from Cycle 1 onwards were converted to ASCII text using the python library {\tt pdftotext}.\footnote{\url{https://github.com/jalan/pdftotext}} The technical justification portion of the proposal that describes and justifies the setup of the observations is removed given it is largely technical in nature and much of the text will appear similar between proposals. The ASCII text for the scientific justification was then combined with the proposal title and proposal abstract stored in the ALMA databases to define the proposal text. Pre-processing steps were applied to the resulting files using {\tt spaCy} \citep{spacy} and custom functions to define and remove stop words, standardize commonly used acronyms, convert British spelling of words to American spelling, and lemmatization. This process created a ``bag of words'' for each proposal. 

Many methods are available to classify documents from the bag of words. We  explored the use of Term Frequency-Inverse Document Frequency (TF-IDF), Latent Semantic Indexing (LSI), and Latent Dirichlet Allocation (LDA). All three methods gave satisfactory and similar results based on how well they can find similar documents which are known \textit{apriori} to share the same scientific category and keywords. We ultimately settled on the LDA method~\citep{blei2003latent} since we found for our application that it was able to match proposals to within the same scientific category at a slightly higher rate than TF-IDF and LSI. 

LDA is an unsupervised machine learning algorithm that assumes a collection of documents can be represented by a mixture of topics, and each topic can be represented by a mixture of words. For a specified number of $N$ topics, the LDA algorithm processes a collection of proposals to define the $N$ topics, and computes a vector for each proposal that contains the relative weight of each topic for that proposal. Thus each proposal $i$ is represented by an $N$-length vector {\it $\vec{p_i}$} that indicates the relative strength of the $N$ topics. The topics were identified using the 17,051 proposals submitted to ALMA starting in Cycle 1 and up to and including Cycle 10. We used the LDA algorithm implemented within the python package \texttt{gensim}. The number of topics is a free parameter that was determined upon iteration by inspecting the words of the topics to judge if the topics were sufficiently distinct based on the knowledge of the type of proposals submitted to ALMA. In addition, we identified the most similar proposals to a given proposal to investigate how well proposals from the same scientific category and keywords can be matched. While subjective, we settled upon $N=50$ topics. 

To gauge the efficacy of our use of the LDA algorithm, we computed the similarity between any two proposals $i$ and $j$ as
\begin{equation}
     S(\vec{p_i},\vec{p_j})=\frac{\vec{p_i} \cdot \vec{p_j}}{||\vec{p_i}||_2\,\, ||\vec{p_j}||_2}.\label{eq:cossim}
\end{equation}
Here, $S(\vec{p_i},\vec{p_j})$ measures the cosine similarity between the topic vectors $i$ and $j$, ranging from 0 (complete dissimilar) to 1 (identical topics). However, the relative values of the similarities are more meaningful than the absolute values. For each proposal, the 10 proposals that have the highest similarity were identified, and we computed the fraction of those proposals that were submitted to the same scientific category as selected by the PI. While this is by no means a perfect measure of the accuracy of the algorithm since there are ambiguities in the definitions of the scientific categories, it does provide a first order indication of how well the algorithm performs since proposals in the same scientific category are expected to have similar content. 

\begin{figure}[ht]
\centering
\includegraphics{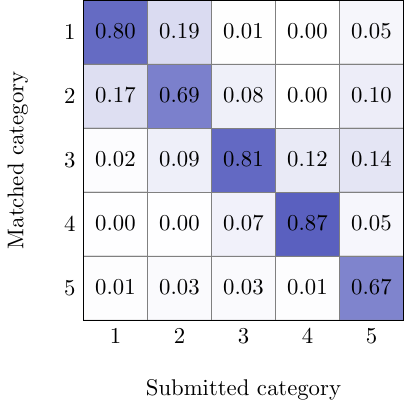}
\caption{Comparison of scientific categories for the closest matching proposals identified using the Latent Dirichlet Allocation model, alongside the category of the submitted proposal. The x-axis represents the categories of the submitted proposals, while the y-axis displays the categories of the ten most similar proposals. Each column is normalized to unity. The results are shown for proposals submitted to ALMA Cycle 10. For each submitted category, the closest matching proposals identified by the machine learning model are found within the same category.}
\label{fig:matrix}
\end{figure}

Figure~\ref{fig:matrix} shows the results for the Cycle 10 proposals. As an example, for proposals submitted to Category 1, 80\% of the matched proposals were also in Category 1,  17\% in Category 2, 2\% in Category 3, 0\% in Category 4, and 1\% in Category 5. The high overlap between Categories 1 and 2 is expected given both contain galaxy proposals and there is no precise division by redshift. More relevant is that only 3\% of the proposals were matched in Categories 3, 4, and 5 combined. Even the 1\% matching in Category 5 is not necessarily spurious, since this category contains transient proposals which overlap with Gamma Ray Bursts in Category 1. Category 5 is the one category that has significant overlap with all other categories. This reflects that large range of topics in this category (e.g., the Sun, transients, pulsars, evolved stars, blackholes), and it also contains the fewest overall proposals, so there may not be 10 proposals that are closely matched. The main conclusion is that while most of closest matched proposals are in the same category as the submitted proposal, there are potential false matches.

\subsection{Reviewer expertise}
\label{subsec:expertise}

In previous applications of machine learning to proposal assignments at astronomical observatories \citep{Kerzendorf19, Kerzendorf20, Strolger23}, the reviewer's expertise has been inferred based on published abstracts or papers available in ADS. This approach has the advantage that it contains an extensive record of a reviewer's publication history. While the publication record of most authors can be inferred accurately from their last name and first initial \citep{Milojevic13}, erroneous matches will inevitably occur. With over 1000 reviewers participating in the review process, it is not practical to perform a manual inspection of the search results to identify any problematic matches. While an Open Researcher and Contributor ID (ORCID) can be used to identify authors uniquely, not all papers are tagged with an ORCID,\footnote{\url{https://orcid.org}} and ALMA does not currently collect ORCIDs from its users. Also, ALMA allows students who are the PI of proposals to serve as reviewers by specifying a mentor, and students may not yet have a publication record.

Instead of using the publication record, we used the proposal history of the reviewers, including the current proposal cycle, to establish their scientific expertise. The inherent assumption is that a reviewer will have expertise in the science topics for proposals in which they are the PI or a co-investigator (coI). The main advantage of this approach is that reviewers are uniquely identified by their ALMA user ID, and, by design, each reviewer is guaranteed to be an investigator on at least one proposal. In addition, each proposal is tagged by the proposal PI with a scientific category and keywords that provide additional information on the topic of the proposal. As discussed below, these keywords can help identify which proposals should be used to determine the reviewer's expertise and to prevent proposal assignments in disparate scientific categories (see Section~\ref{subsec:assignments}). A downside to using the proposal text is that the proposal history will not capture the full expertise of a reviewer, especially for senior astronomers who are new users to ALMA and may have a substantial publication history.   

Each reviewer is associated with one or more proposals as either PI or coI. The topical vector for each proposal associated with a reviewer is determined as described in Section~\ref{subsec:proptopics}. These topical vectors are averaged to determine the reviewer's overall expertise ($\vec{r}$), which is also an $N$ length vector. Similar to Equation~\ref{eq:cossim}, the cosine similarity between the $i^{th}$ proposal and the $k^{th}$ reviewer's expertise is used to assess where the reviewer is suitable for evaluating that proposal:
\begin{equation}
     S(\vec{p_i},\vec{r_k})=\frac{\vec{p_i} \cdot \vec{r_k}}{||\vec{p_i}||_2\,\, ||\vec{r_k}||_2}.\label{eq:cosrev}
\end{equation}
The similarity score $S(\vec{p_i},\vec{r_k})$ ranges from 0 to 1, where 0 indicates no alignment between the reviewer's expertise and the proposal topic, and 1 indicates a perfect match. While the absolute values of the similarity scores may be less critical, their relative values are meaningful for comparing how well each proposal aligns with a reviewer’s expertise.

This general approach works well for most reviewers, but the implementation also accounted for those with expertise or proposal submissions in multiple categories. Special considerations were also made for categories and keywords with limited reviewers available. Appendix~\ref{app:revexp} provides further details on how reviewer expertise was calculated to address these specific situations.

\subsection{Proposal assignments}
\label{subsec:assignments}

Given the formalism to compute the similarities between the proposal topics and reviewer's expertise as defined in Section~\ref{subsec:expertise}, we proceed to assign proposals to reviewers. The goal is for each proposal to be reviewed 10 times and for each reviewer to be assigned 10 proposals. The main input is a matrix that contains similarities between the reviewer expertise and all submitted proposals for a given cycle. The rows in the matrix correspond to each submitted proposal, and the columns in the matrix correspond to each submitted proposal for the cycle. If an individual is reviewing on behalf of more than one proposal, they will appear as separate rows in the matrix. 

Before assigning proposals, similarities are set to a negative number if a proposal assignment is not permitted. There are two main reasons for restricting proposal assignments in this manner. First, potential assignments are flagged if the reviewer has a known conflict of interest on the proposal. Conflicts of interest are defined if the reviewer is a PI or coI on the potential proposal assignment proposal, the PI is deemed a close collaborator based on the proposal history \citep[see][]{DonovanMeyer22}, or the reviewer identified a PI or coI as a close collaborator. In these cases, the similarities are set to a value of $-100$, which effectively prohibits the assignment of the conflicted proposal. The second case is if the potential assigned proposal is from a distinct scientific category compared to the reviewer's expertise (see Section~\ref{subsec:expertise}). 
While such proposals will typically have low similarities, this additional measure reduces the risk of a false positive match. In practice, the new similarity is set as $-1.01 + S(\vec{p_i}, \vec{r_k})$. While a negative number does not preclude a proposal assignment, it makes it less likely and can be checked after the assignments are made.

Once the similarity matrix is established, proposals are assigned to reviewers. The goal is to define a metric that measures the success of these assignments and optimize it during the assignment process. One common approach is to maximize the total similarity scores between proposals and reviewers, which increases overall expertise in the review process. However, this method may lead to unfairness if certain types of proposals are disproportionately assigned to less-qualified reviewers. ALMA instead adopted the approach used in PeerReview4All, which optimizes a ``leximin fairness'' metric   \citep[code available at \url{https://github.com/niharshah/peerreview4all}]{Stelmakh21}. This method optimizes the similarity for the most ``disadvantaged'' proposals; i.e., proposals that have the least overall expertise among  reviewers. Once the assignment for the most disadvantaged proposal is optimized, the algorithm moves to the next most disadvantaged, and so on. Proposals in fields with many qualified reviewers are assigned last, as they have more reviewer options available. The metric used by PeerReview4All is solved using the \texttt{gurobi} python optimizer \citep{gurobi}.

The PeerReview4All code was adapted to handle special cases. Student reviewers specify a mentor, who may also serve as a reviewer for other proposals. To ensure fairness, we required that the proposal sets for students and mentors be disjoint. After each round of assignments, the similarity matrix was updated to mark as conflicted any proposals previously assigned to a reviewer or their mentor. In addition, conflicts were updated for reviewers assigned multiple sets, to ensure that they were not assigned the same proposal twice.

To prevent ``niche" proposals from being repeatedly reviewed by the same group, further restrictions were imposed on the number of proposals that a reviewer could be assigned within the same topic. These restrictions were specifically applied to proposals related to Very Long Baseline Interferometry, Target of Opportunity, and solar studies.

\section{Results}
\label{sec:evaluation}

Using the framework described in Section~\ref{sec:implementation} to compute the similarity between proposals and reviewer expertise, we now assess whether the new process improved proposal assignments compared to previous cycles. The new assignment approach was implemented in ALMA Cycle 10, and we evaluate its effectiveness by comparing it to Cycles 8 and 9. To ensure a uniform comparison, we computed similarities for all three cycles using the same LDA model trained on the same set of proposals.

\subsection{Evaluating similarity computation on Cycle 8 and Cycle 9 data}
\label{subsec:test89}

We evaluated the effectiveness of the similarity scores (calculated as described in Section~\ref{subsec:expertise}) by using data from previous review cycles. Specifically, we examine whether the computed similarities between a reviewer’s expertise and the proposal topics accurately reflect the reviewer’s expertise. During the distributed peer review process used by ALMA in Cycles 8 and 9, reviewers were asked to self-assess their expertise on each assigned proposal by indicating if the assigned proposal is in their field of expertise (i.e., the review considers themselves an ``expert''), they are somewhat knowledgeable, or they have little to no knowledge of the topic. If our similarity scores are a reliable measure of expertise, we would expect that reviewers who self-identify as experts in a proposal topic will, on average, have higher similarity scores compared to those who report having little or no knowledge of the topic.

\begin{figure}[ht]
\centering
\includegraphics{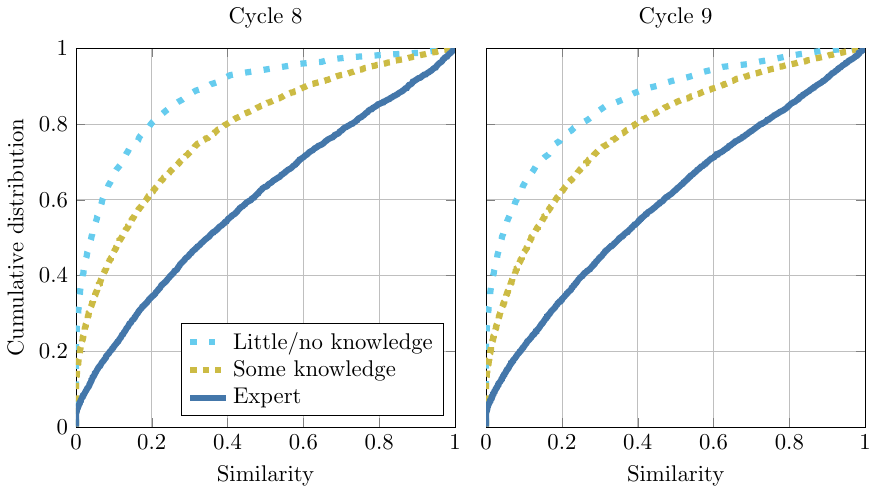}
\caption{
  \label{fig:cycle8_9_survey}
  Cumulative distributions of the similarities between the reviewer's expertise and the topic of the assigned proposal in Cycle 8 (left) and Cycle 9 (right), when categories and keywords were used to assign proposals. The similarities were computed retroactively to evaluate the machine learning algorithm. The results are shown for proposal assignments where reviewers indicated they are experts (solid curve), had some knowledge of the proposal (dashed), and had little/no knowledge of the proposal (dotted). Similarities tend to be higher when reviewers indicate expertise in the proposal, suggesting that these similarities have predictive value for assessing the suitability of reviewer assignments.
}
\end{figure}

Figure~\ref{fig:cycle8_9_survey} shows the similarities of the proposal assignments in Cycle 8 and 9 based on the reviewer's self-declared expertise. In these cycles, assignments were made solely using categories and keywords, without considering similarity metrics.  When reviewers identified a proposal as being within their area of expertise, the distribution of similarities has a median of approximately 0.35 for both cycles. In contrast, when reviewers indicated only partial familiarity with a proposal, the median similarity dropped to around 0.12, and for those with little to no expertise, the median similarity was 0.04. 

As expected, the highest median similarity occurred when reviewers declared expertise, and the lowest when they indicated little or no knowledge, with comparable trends across Cycles 8 and 9. However, similarity alone is not a perfect indicator of expertise. Some assignments had near-zero similarity scores, yet reviewers reported they were experts, while other proposals had similarity scores close to 1, but reviewers expressed limited or no knowledge of the subject. This may point to both the inherent imprecision in survey data and the limitations of assessing expertise based solely on proposal history. Despite these limitations, the correlation between self-declared expertise and similarity suggests that these metrics can help improve the alignment of proposals with a reviewer's scientific background.

\subsection{Comparing similarities for Cycle 9 and Cycle 10 assignments}
\label{subsec:cycle10}

In this section, we analyze the role of the optimization procedure for computing the assignments (Section~\ref{subsec:assignments}) given the computed similarities. In Cycle 9, proposals were assigned to reviewers based on the proposal category and keywords \citep{DonovanMeyer22}; i.e., the ``old'' algorithm. In Cycle 10, we used the similarities and assignment algorithm described in Section~\ref{sec:implementation}; i.e., the ``new'' algorithm. In both cycles, a small percentage of initial assignments was reassigned after the reviewers declared conflicts of interest.\footnote{The percentage of assignments where reviewers declared a conflict of interest increased in Cycle 10 to 2.4\% from 1.4\% in Cycle 9, as more reviewers declared the assignments conflicted scientifically with their own proposal when the machine learning algorithm was adopted.}

\begin{figure}[ht]
\centering
\includegraphics{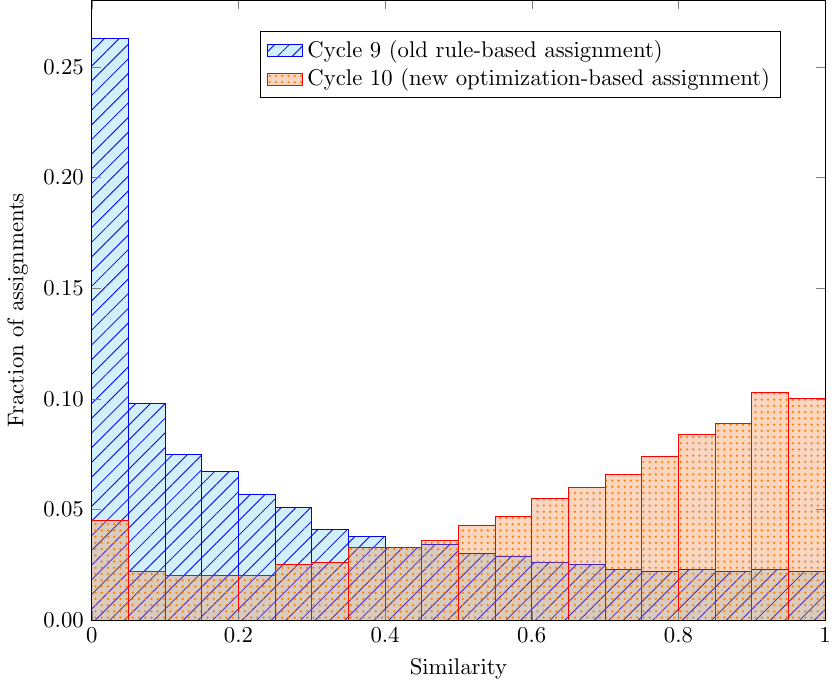}
\caption{
  \label{fig:cycle10_sim}
  Histogram of the similarities between the reviewer expertise and the assigned proposal for Cycle 9 under the old algorithm (blue, hatched histogram), and in Cycle 10 for the new algorithm (orange, dotted histogram). With the new assignment process adopted in Cycle 10, 
  the median similarity increased to 0.71 from 0.20 in Cycle 9, indicating a higher level of expertise in the proposal assignments was achieved with the new assignment process (see Section~\ref{subsec:assignments}).
}
\end{figure}

Figure~\ref{fig:cycle10_sim} shows a histogram of the similarity scores for the actual proposal assignments in Cycle 9 (blue histogram) and Cycle 10 (orange histogram), where the latter used the assignment optimization method described in Section~\ref{subsec:assignments}. We find a substantial increase in the similarity between the reviewer's expertise and the assigned proposals with the new algorithm, as the median similarity increased from 0.20 in Cycle 9 to 0.71 in Cycle 10.

In principle, the increase in similarity scores could be attributed to demographic changes in the reviewer or proposal pools, but this seems unlikely. In Cycle 9, 33\% of the reviewers were new compared to Cycle 8, yet the median similarity score remained nearly unchanged (0.18 in Cycle 8 vs. 0.19 in Cycle 9; see Figure~\ref{fig:cycle8_9_survey}). Similarly, in Cycle 10, the proportion of new reviewers was again 33\% relative to Cycle 9, but the median similarity score increased to 0.71. Given that reviewer turnover remained consistent across cycles and that the increase in similarity coincided with a change in the assignment algorithm, it is more plausible that the improvement reflects an improved alignment between reviewer expertise and proposal topics, rather than demographic changes.

\subsection{Cycle 10 reviewer survey}
\label{subsec:survey}

After completing their reviews, reviewers had the option to complete a survey to self-assess their expertise on their assigned proposals. While self-reported expertise is subjective and may vary with individual perceptions of ``expertise'' \citep[see, e.g.,][]{Karpen18}, the survey offers an indication of the suitability of the review assignments. More importantly, trends across cycles reveal how reviewers perceive the quality of their assignments, assuming minimal changes in reviewer demographics between cycles.

\begin{figure}[ht]
\centering
\includegraphics{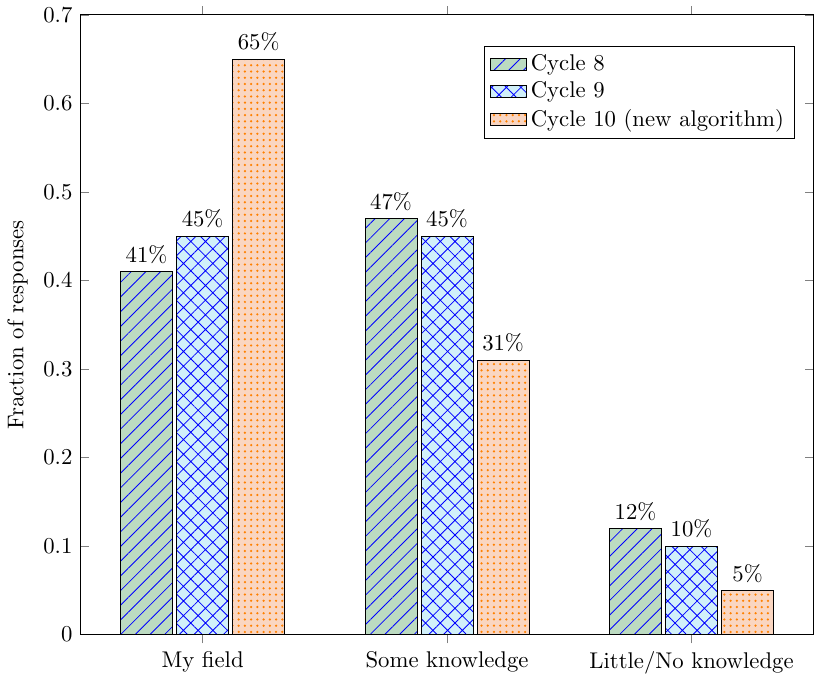}
\caption{
  \label{fig:rsurvey}
  Histogram of the reviewer's self-expertise on their proposal assignments in Cycles 8--10. With the new assignment algorithm implemented in Cycle 10, the percentage of reviewers reporting expertise in their assigned proposals increased by 20 percentage points compared to the previous cycle, while the percentage of those declaring little or no knowledge was reduced by half.
}
\end{figure}

Figure~\ref{fig:rsurvey} shows a histogram of survey results from Cycles 8, 9, and 10. Following the implementation of the new assignment process in Cycle 10, the percentage of reviewers identifying as experts on their assigned proposals increased from 45\% in Cycle 9 to 65\% in Cycle 10. Simultaneously, the proportion of assignments where reviewers reported little or no knowledge of the proposals dropped from 10\% in Cycle 9 to 5\% in Cycle 10. These results indicate that the new assignment algorithm achieved a better overall match between reviewers’ expertise and their assigned proposals.

\subsection{Topological accuracy of the proposal rankings}
\label{subsec:top}

In this section, we evaluate a key tradeoff in expertise-based reviewer assignments: balancing the expertise of assigned reviewers with the topology of the assignments. Strategies to assign proposals to reviewers typically prioritize maximizing relevant scientific expertise. However, strictly focusing on expertise can lead to the isolation of certain proposal groups, especially those within the same subtopic. For example, imagine a scenario where half of the reviewers are exclusively assigned to review one group of proposals, while the other half reviews the rest. This division hampers direct comparison between the two groups, limiting the ability to rank proposals effectively across groups.

To quantify how assignment topology affects ranking accuracy, we use the methodology from \citet{shah2016estimation} to assess how the structure of reviewer assignments impacts the final outcomes, assuming all reviewers have equal expertise. Consider a graph where each proposal is a vertex. An edge is placed between two vertices if there is at least one reviewer who reviews both proposals (that is, if the two proposals are directly compared). The weight of an edge is the number of times that pair of proposals is assigned to the same proposal set. The Laplacian for this graph determines the topology of the proposal assignments. Following \citet[see also~\citealp{hendrickx2020minimax,lee2021minimax}]{shah2016estimation}, 
the trace of the (pseudo) inverse of the Laplacian is proportional to the ranking accuracy.

To estimate the relative accuracy of the actual assignments in Cycle 10, we set the error in the actual assignments to unity. We then ran simulations where the reviewers were assigned proposals at random across any scientific category, subject to the constraint that each reviewer is assigned 10 proposals and each proposal is assigned to 10 reviewers. The random assignment strategy is known to be optimal~\citep{shah2016estimation} in terms of topology when all reviewers have identical expertise. We find that the relative uncertainty is 0.86 for the random assignment. Thus, optimizing assignments for expertise increases the uncertainty in final outcomes by only 14\% based on the topology of the proposal assignments.

\section{Summary and conclusions}
\label{sec:summary}

The ALMA peer review system has undergone significant evolution in response to the increasing volume of submitted proposals. First, ALMA adopted a distributed peer review system to streamline the logistics of reviewer identification \citep{DonovanMeyer22}. This paper presents the implementation and outcomes of the next major advancement: the integration of machine learning \citep{blei2003latent} and optimization techniques \citep{Stelmakh21} to assign proposals to reviewers.

The assignment process begins by predicting each reviewer’s expertise for every submitted proposal. This is achieved through topic modeling using machine learning techniques, which analyze the topics of submitted proposals and infer reviewers' expertise based on their proposal submission history. We calculate the similarity between a reviewer's expertise and the topics of each proposal, using this metric as a measure of the reviewer's suitability. To further enhance the accuracy of assignments and address potential mismatches in the machine learning predictions, the process incorporates keywords provided by proposal teams alongside expertise keywords specified by reviewers, ensuring more robust and precise reviewer-proposal matches.

The second step in the assignment process employs optimization algorithms that are used in computer science \citep{Stelmakh21}.  The optimization algorithms assign proposals to reviewers by prioritizing the most ``disadvantaged” proposals (i.e., those with the lowest similarity scores and least available reviewer expertise), thereby improving the overall quality of matches.

Various evaluation metrics highlight the effectiveness of the new algorithm. Testing the similarity computations on proposal assignments from Cycles 8 and 9, which were based solely on keywords, reveals a strong correlation between the similarity scores and reviewers' self-reported expertise, demonstrating the predictive value of these scores. After applying the new assignment algorithm in Cycle 10, the median similarity between proposal topics and reviewer expertise increased significantly, from 0.20 in Cycle 9 (under the previous algorithm) to 0.71 in Cycle 10. This improvement reflects a substantial enhancement in the alignment of expertise with assignments. The reviewer survey validated these results, showing that the proportion of assignments where reviewers reported high expertise rose from 45\% in Cycle 9 to 65\% in Cycle 10, while assignments with little or no expertise dropped from 10\% to 5\%. Moreover, the new system completely eliminated the need for reassignment due to gross mismatches, saving three to five days of manual effort per cycle.

Despite these successes, challenges remain. Since a reviewer's expertise is inferred from their proposal history, machine learning algorithms can be quite effective in assigning proposals that are similar to a reviewer's own submitted proposal. This raises questions of potential conflicts of interest (which may or may not be flagged by the reviewer). For instance, in competitive reviewing scenarios, reviewers may downrate competing papers to favor their own~\citep{Merrifeld09, balietti2016peer}, which may be mitigated using optimization methods to prevent such outcomes~\citep{alon2011sum,dhull2022price} or detecting such behavior~\citep{stelmakh2020catch}. Competitive reviewing can also incentivize collusion rings where reviewers manipulate assignment systems to support papers of colluding researchers in exchange for favors~\citep{littman2021collusion,Vijaykumar2020Architecture,hsieh2024vulnerability}. Randomization in assignments can mitigate such risks, though not entirely~\citep{jecmen2020manipulation}. 
Addressing these challenges remains an ongoing effort.

While currently employed automated assignment systems maximize each assigned reviewers' expertise for the papers, it raises questions of whether reviewers should be assigned such a narrow range of topics or if they should be assigned more diverse proposals~\citep{goyal2024causal}. In addition, the topology of proposal assignments impacts the accuracy of the final rankings, as discussed in Section~\ref{subsec:top}. Future work should explore these factors to identify optimal trade-offs.

Our experience also suggests that the implementation of machine learning could be improved. In a small fraction of cases, proposals were matched to reviewers based on words that are relatively rare, but can appear across different scientific categories (see Section~\ref{subsec:expertise}). For example, a proposal to observe polarization in the Sun will have a finite similarity to a proposal to observe dust polarization in high-redshift galaxies since both proposals use the term ``polarization.'' In the current deployment, we use the keywords provided by the PIs and the reviewers to mitigate such assignments and increase the robustness of the overall process. Some of these mismatches could likely be reduced with improved machine learning algorithms (e.g., using transformer-based language models and/or fine-tuning to ALMA-relevant corpora). However, it is likely that some level of mismatch will persist~\citep{stelmakh2023gold}, as inherent limitations and uncertainties are a natural part of such automated processes.

The success of these methods in ALMA’s peer review process highlights the broader potential of machine learning and optimization for addressing challenges in scientific peer review, paving the way for its wider adoption across the research community.

\section*{Acknowledgements}

We thank Victoria Cat\'an and Nicolas Buzeta for their help in developing the initial topic modeling algorithms, and Ignacio Toledo for his assistance in using the Dataiku platform. The work of NBS was supported by NSF CAREER 1942124.

\software{
\texttt{gensim} \citep{Gensim}, 
\texttt{gurobi} \citep{gurobi},
\texttt{PeerReview4All} \citep{Stelmakh21}, 
\texttt{spaCy} \citep{spacy}
}

\bibliography{references}

\appendix 

\section{Identifying A Reviewer's Expertise}
\label{app:revexp}

Section~\ref{subsec:expertise} presented the general approach to identify the expertise of each reviewer. Although this basic method works well for most reviewers, several practical issues can arise. Some reviewers are involved in proposals across distinct topics, either as PIs or coIs. Combining all their proposals into a single expertise vector may dilute their specific areas of expertise, resulting in decreased similarity when comparing against proposals focused on only one topic. Additionally, some reviewers claim expertise in a broad range of topics and submit proposals across multiple categories. Averaging their expertise based on all submitted proposals may lead to a dominance of the category that contains the most proposals. Moreover, some reviewers assert expertise in categories that differ from those of their submitted proposals, indicating a preference for reviewing in certain topics. Assuming the reviewers' declared expertise at face value and including all proposals blindly could result in assignments outside their actual areas of expertise. Finally, there can be false positives arising from the LDA topic modeling, where a reviewer's expertise may appear similar to a proposal topic, but upon closer inspection, the proposal may not be relevant to the reviewer's expertise. Such false matches can occur due to common words or phrases shared across multiple categories, such as ``masers'' and ``polarization.'' This suggests that the topics identified in the LDA model may require further refinement.

To address these challenges, we utilize the keywords associated with a reviewer’s expertise to select the proposals that will be used to infer their expertise. The overall goal is to evaluate a reviewer’s expertise based on proposals that are as closely aligned as possible with the topic of the parent proposal, while also respecting the reviewer’s self-declared areas of expertise. Each ALMA proposal is submitted to a specific scientific category, defined as the {\em parent} category. We have predefined {\em similar} categories that significantly overlap with the parent category in terms of scientific content (see Figure~\ref{fig:matrix}). For instance, Categories 1 and 2 are considered similar because they both pertain to galaxies, while Categories 3 and 4 are similar due to their focus on the formation of stars and circumstellar disks. In contrast, Categories 1 and 2 are distinct from Categories 3, 4, and 5, and Categories 3 and 4 are distinct from Categories 1, 2, and 5. Category 5 is viewed as distinct from all other categories due to its diverse set of keywords.

With these definitions in place, a reviewer’s expertise is constructed based on whether they have provided their expertise keywords and whether they are the PI or coI on proposals. Table~\ref{tbl:expertise} outlines the sequence of steps used to infer the reviewer’s expertise. In the first step, if a reviewer has declared expertise on any keywords in a parent category C and has served as the PI on proposals in that category during any cycle, the expertise vector is constructed from the proposals in Category C where they are the PI. If the conditions of Step 1 are not met, Step 2 checks whether the reviewer has expertise in a similar category and has served as the PI on proposals in that category. If those conditions are also not met, Step 3 checks if the reviewer has expertise in a distinct proposal category and has served as the PI on proposals in those distinct categories. If no suitable proposals are found in these steps, the process is repeated by checking proposals in which the reviewer is a coI (Steps 4--6).

If a reviewer has not provided their expertise, Steps 7--9 examine categories in which they have submitted proposals as the PI, followed by Step 10, which considers categories where they have served as a coI in the parent category. Finally, if a reviewer has declared their expertise but has neither been a PI nor a coI on any proposals within those expertise categories, a constant similarity value is assigned for proposals in their declared expertise category.

An additional subtlety to this procedure is that there are some scientific categories (e.g., Category 5) and subtopics within categories (e.g., Solar System within Category 4) where there are a limited number of proposals and therefore a limited number of available reviewers. The above procedure could reduce the chance reviewers are assigned such proposals if they submitted proposals in other topics. If a reviewer declared they had expertise in these less popular keywords and has been PI or coI on such proposals, separate expertise vectors were computed for the parent proposal category and these other topics. The appropriate expertise vector is used when computing the similarities depending on the keyword of the proposal.

The last column in Table~\ref{tbl:expertise} indicates the percentage of expertise vectors that were used in each step in Cycle 10. In the vast majority of cases (94.3\%), the expertise vector was computed based on the same category in which the reviewer submitted a proposal (Steps 1 and 7 combined). The remaining cases mitigated situations where the reviewer has not been a PI of proposals in the parent category.

\begin{deluxetable}{cccccr}
\label{tbl:expertise}
\tablecaption{Steps to construct the reviewer's expertise}
\tablehead{
\colhead{Step} & 
\colhead{Provided} &
\colhead{Parent} &
\colhead{Similar} &
\colhead{Distinct} &
\colhead{Percentage}\\
\colhead{} & 
\colhead{Expertise} &
\colhead{category} &
\colhead{category} &
\colhead{category}
}
\startdata
 1 & \checkmark & PI  &     &   &91.0\%\\
 2 & \checkmark &     & PI  &   &1.7\%\\
 3 & \checkmark &     &     &PI &0.9\%\\
 4 & \checkmark & coI &     &   &2.4\%\\
 5 & \checkmark &     & coI &   &0.1\%\\
 6 & \checkmark &     &     &coI&0.2\%\\
 7 &            & PI  &     &   &3.3\%\\
 8 &            &     & PI  &   &0.1\%\\
 9 &            &     &     &PI &0\%\\
10 &            & coI &     &   &0.1\%\\
11 & \checkmark &     &     &   &0.4\%\\
\enddata
\end{deluxetable}
\end{document}